\documentclass[prl,twocolumn]{revtex4}
\usepackage{graphicx, epsfig}
\begin{document}


\title{Mechanism for modulated structures in Ni-Mn-Ga: An EXAFS Study}

\author{P. A. Bhobe, K. R. Priolkar, P. R. Sarode}
\affiliation{Department of Physics, Goa University, Goa, 403 206 India.}

\date{\today}

\begin{abstract}
The local atomic structure of Ni-Mn-Ga alloys was explored using Mn and Ga K-edge extended X-ray fine structure (EXAFS) measurement. The changes occuring in the L1$_0$ sub-cell of the martensitic unit cell and the bond lengths obtained from the analysis enables us to propose a scheme for the structural distortions responsible for the modulations and shuffling of the atomic planes in the martensitic phase of Ni-Mn-Ga system. The EXAFS analysis also suggests the changes in hybridization of Ga-$p$ and Ni-$d$ orbitals associated with the local symmetry breaking upon undergoing martensitic transition.     
\end{abstract}
\maketitle

Martnesitic transformations are first-order, diffusionless, displacive, solid-solid phase transformation taking place upon cooling below a characterisitc temperature T$_M$ from a high symmetry initial phase to a low-symmetry structure \cite{nisi}. Extensive studies in the past have attributed the structural transformations to the phonon anamolies occuring in the parent phase \cite{shap1, shap2}. Here, an incomplete softening of the [$\zeta$ $\zeta$ 0]TA$_2$ phonon mode at a particular wave vector $\zeta_0$(corresponding to the periodicity of the martensitic phase) with displacement along the [110] direction takes place. Such a phonon softening is belived to be due to contribution from electron-lattice coupling and nesting of the Fermi surface \cite{zhao, dug}. 

Ni$_2$MnGa is one of the shape memory alloys that exhibits martensitic transformation upon cooling through 220K. Moreover, it is ferromagnetic with a Curie temperature, T$_C \sim$ 370K making it a technologically important {\it magnetic} shape memory alloy. Inelastic scattering measurements made on the single crystals showed that the softening of the [$\zeta$ $\zeta$ 0]TA$_2$ phonon mode occurs at wave vector $\zeta_0$$\sim$ 0.33 and has been related to the specific shape of the Fermi surface of Ni$_2$MnGa \cite{Zheu}. Alot of theoretical and experimental study that reveal different properties related to martensitic transformation have been carried out \cite{ayu, vasil1}. The early neutron diffraction study \cite{Web} determined the structure of the stoichiometric Ni$_2$MnGa to be cubic L2$_1$ Heusler type with a = 5.825\AA. Many other investigations have been carried out on the near-stiochiometric alloys confirming that the martensitic structure is a tetragonal distortion of the initial cubic lattice. The low temperature crystal structure of the non-stoichiometric Ni-Mn-Ga alloys revealed that there exists different intermartensitic transformations as the lattice is subjected to periodic shuffling of the (110) planes along the [1$\bar 1$0]$_P$ direction of the initial cubic system \cite{mart} with modulation period dependent on the composition as summarized in \cite{pons}. Recent calulations by \cite{zay1, zay-role} indicate the importance of modulated structure and the shuffling of atomic planes in stabilizing the martensitic structure. In spite of intense efforts, the underlying mechanism giving rise to such a phase transformation is still not well understood. The nature of modulations forming the super structures and the driving force for the martensitic transformation in these alloys is currently at debate. An understanding at the microscopic level of such transformation can be achived by making a comparative study of the unit cell in going from austinitic to martensitic phase. Thus a precise knowledge of the changes occuring in the L1$_0$ sub-unit of the martensitic phase is fundamental in understanding the mechanism involved in martensitic tansformations. It is with this objective that the present experiment was undertaken.

In this letter we report EXAFS studies at Mn and Ga K-edge in the Ni-Mn-Ga system to explore the changes in local environment around these metal ions in the austenitic and martensitic phase. We have carried out the measurements on two alloy compositions: Ni$_2$MnGa with T$_M$ = 220K and Ni$_{2.16}$Mn$_{0.84}$Ga with T$_M$ = 316K. Therefore, at room temperature, Ni$_2$MnGa and Ni$_{2.16}$Mn$_{0.84}$Ga represents the austinitic and martensitic phases of the Ni-Mn-Ga system respectively. It may be noted that the the sensitivity of XAFS analysis to distinguish between Mn (Z = 25) and Ni (Z = 28) occupying the same crystallographic site is limited. Hence as far as the XAFS analysis is concerned, the off-stoichiometric composition is not significantly different from the stoichiometric one. Moreover, in Ni$_{2.16}$Mn$_{0.84}$Ga the excess Ni substituting at Mn site does not replace even one full atom of Mn in a unit cell.

Polycrystalline ingots of Ni$_2$MnGa and Ni$_{2.16}$Mn$_{0.84}$Ga were prepared by conventional arc-melting method\cite{vasil2, pab}. Room temperature powder X-ray diffraction patterns recorded on Rigaku D-MAX IIC diffractometer with Cu K$\alpha$ radiation indicated the samples to be phase pure and Energy dispersive X-ray (EDX) analysis confirmed the compositions to be nominal.  

Absorbers for the EXAFS experiments were made by spreading very fine powder on a scotch tape avoiding any sort of sample inhomogentiy and pin holes. Small strips of the sample coated tape were cut and were held one on top of other. Enough number of such strips were adjusted to give absorption edge jump ($\Delta\mu x$) of about 1. Room temperature EXAFS at Mn and Ga K-edges were recorded in the transmission mode at the EXAFS-1 beamline at ELETTRA Synchrotron Source using Si(111) as monochromator. The incident and transmitted photon energies were simultaneously recorded using gas-ionization chambers filled with mixtures of He-N$_2$ for Mn edge and Ar-N$_2$ for Ga edge. Measurements were carried out from 300eV below the edge energy to 1000eV above it with a 5eV step in the pre-egde region and 2.5eV step in the EXAFS region. At each edge, three scans were collected for each sample. Data analysis was carried out using IFEFFIT suite wherein theoretical fitting standards were computed with ATOMS and FEFF6 programs \cite{rav, zab} and fitting was done using FEFFIT program \cite{new}.
 
\begin{figure}[h]
\epsfig{file=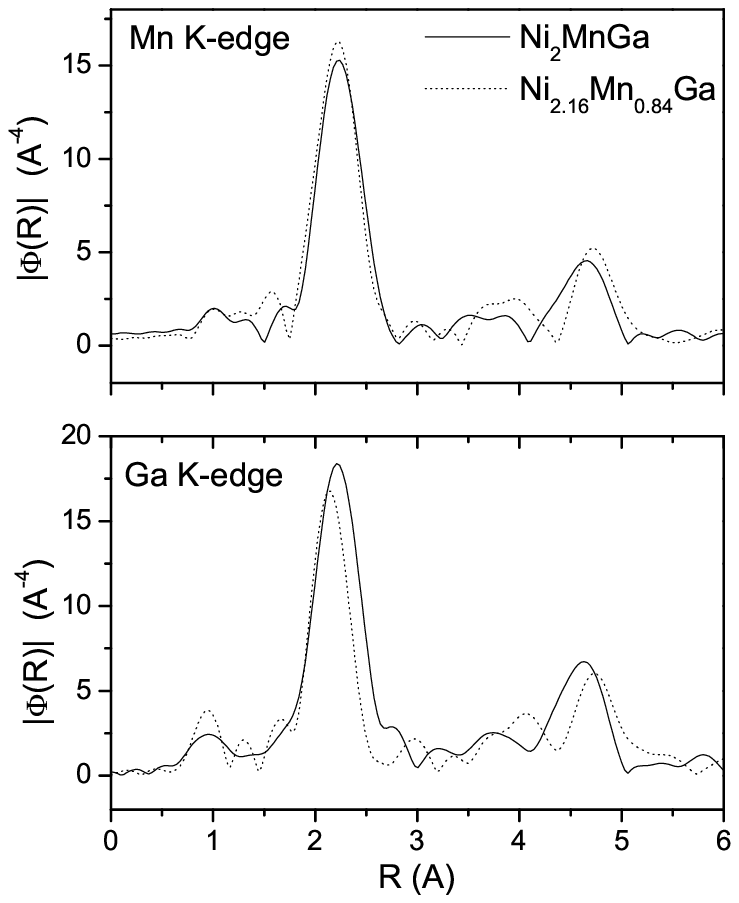, width=7cm, height=8cm}
\caption{\label{xafs-all} The $k^3$ weighted FT EXAFS spectra for Mn and Ga K-edge for Ni$_2$MnGa and Ni$_{2.16}$Mn$_{0.84}$Ga. An indication of change in the nearest neighbour bond-length for Ni$_{2.16}$Mn$_{0.84}$Ga is evident from the shift of the first peak ($\sim$2.5\AA) in Ga K-edge spectra.}
\end{figure}

Magnitude of $k^3$ weighted Fourier transform (FT) spectra for Mn and Ga K-edge EXAFS in austenitic Ni$_2$MnGa and martensitic Ni$_{2.16}$Mn$_{0.84}$Ga are presented in Fig.\ref{xafs-all}. The first peak at around R = 2.5\AA~ is due to scattering from the nearest-neighbour shell comprising of 8 Ni atoms. Martensitic transition being diffusionless, no drastic variation atleast in the first shell is expected in the two phases. Indeed, it is seen from Fig.\ref{xafs-all} that the first peak position remains unchanged in the FT spectra of Mn K EXAFS for the two samples. However, in case of FT of Ga K EXAFS a shift to lower R in the position of the first peak is observed for the sample in martensitic phase. This is very important observation in context to modulated structures in martensitic phase. Further, the broad peaks in the range R = 2.8 - 5.0\AA~ are due to the combined contribution from the second to fourth single scattering paths and some relatively weak multiple scattering paths. In this region of R, a difference in spectral signatures of the two alloys is quite evident. This can be attritubed to the lowering of symmetry from the parent cubic structure due to martensitic transition. 

\begin{figure}[h]
\epsfig{file=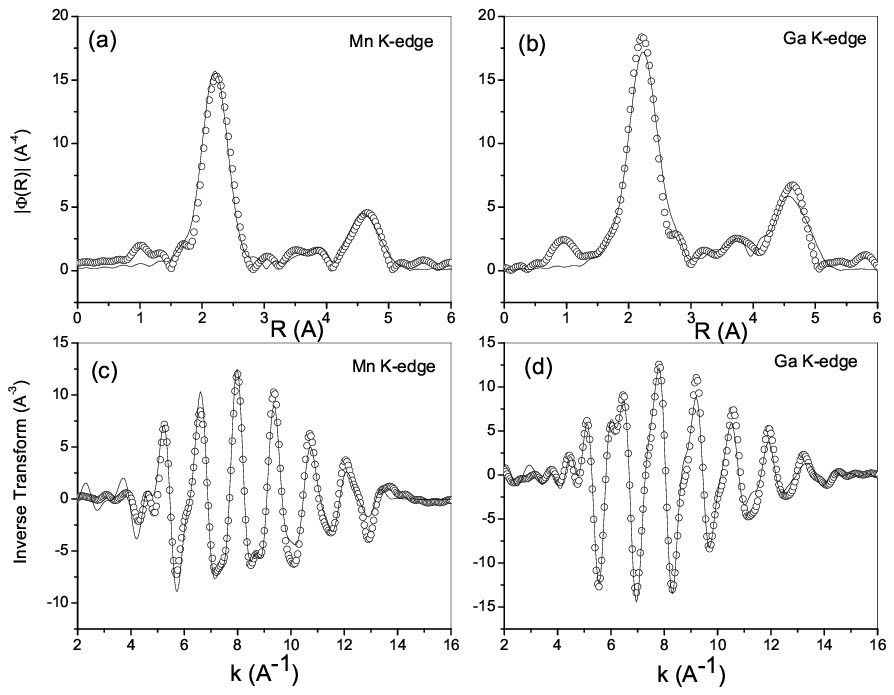, width=7cm, height=8cm}
\caption{\label{Ni200-xafs} Fit results for Mn and Ga K-edge EXAFS in Ni$_2$MnGa.The upper panel show the fit in R space and the lower panel show the $k^3$ weighted back transformed spectra. The circles represent the data and the solid line is the fit.}
\end{figure}

EXAFS spectrum of Ni$_2$MnGa for both the edges was fitted using the common set of variable parameters with Fm3m space group and lattice constant 5.825\AA. In this model, the correction to the path lengths was refined with a constraint, 
$$ \delta R = \delta r_1 \times \frac{R_{eff}}{R_{nn1}}$$ 
where $R_{nn1}$ is the nearest neighbour distance, kept fixed to 2.5223\AA~ obtained from the lattice constant, $R_{eff}$ is the calculated bond length obtained from FEFF and $\delta r_1$ is the change in first neighbour distance. This approach reduces the number of variable parameters in the fit. The thermal mean-square variation in the bond lengths, $\sigma^2$, were varied independently for each kind of single scattering paths considered in the fit. The fitting was carried out in $R$-space in the range 1\AA~ to 5\AA. The final fitted parameters obtained are presented in the Table\ref{tab1} and the fittings in the $R$ space and the back transformed $k$-space are shown in Fig.\ref{Ni200-xafs}. As seen from the figure, the fitting is quite satisfactory.

\begin{table}[h]
\caption{Results of the fits to the Mn and Ga edge data of Ni$_2$MnGa. Figures in parantheses indicate uncertainity in the last digit.}
\vspace{0.2cm}
  \centering
  \begin{tabular}{lccc}
Mn K-edge & $k$ range: (2- 15)\AA$^{-1}$ & $k$-weight: 3 & $R$ range: (1-5)\AA\\
\hline
\hline
Bond Type & Bond Length  & $\sigma^2$ & Coordination \\
& R (\AA) & (\AA$^2)$ & Number\\
\hline
Mn-Ni1 & 2.523(3) & 0.0093(4) & 8 \\  
Mn-Ga1 & 2.914(3) & 0.035(2) & 6 \\
Mn-Mn1 & 4.12(3) & 0.024(6) & 12 \\
Mn-Ni2 & 4.83(3) & 0.022(4) & 24 \\
\hline
\hline 
&&&\\

Ga K-edge & $k$ range: (2- 15)\AA$^{-1}$ & $k$-weight: 3 & $R$ range: (1-5)\AA\\
\hline
\hline
Bond Type & Bond Length  & $\sigma^2$ & Coordination \\
& R (\AA) & (\AA$^2$) & Number\\
\hline
Ga-Ni1 & 2.521(2) & 0.0079(3) & 8\\
Ga-Mn1 & 2.911(2) & 0.029(6) & 6\\
Ga-Ga1 & 4.12(2) & 0.021(4) & 12 \\
Ga-Ni2 & 4.83(2) & 0.016(2) & 24 \\
\hline
\hline 
 \end{tabular}
    \label{tab1}
\end{table}

\begin{table}[h]
  \caption{Results of the fits to the Mn and Ga edge data of Ni$_{2.16}$Mn$_{0.84}$Ga. Figures in parantheses indicate uncertainity in the last digit.}
  \centering
  \begin{tabular}{lccc}
Mn K-edge & $k$ range: (2- 15)\AA$^{-1}$ &  $k$-weight: 3 & $R$ range: (1-5)\AA\\
\hline
\hline
Bond Type & Bond Length  & $\sigma^2$ & Coordination \\
& R (\AA) & (\AA$^2$) & Number\\
\hline
Mn-Ni1 & 2.532(2)  & 0.0083(3) & 8 \\  
Mn-Ga1 & 2.755(6) & 0.0088(6) & 4 \\
Mn-Ga2 & 3.24(6)  & 0.017(7) & 2 \\
Mn-Mn1 & 3.94(3)  & 0.018(6) & 4 \\
Mn-Mn2 & 4.25(1)  & 0.013(2) & 2 \\
Mn-Ni2 & 4.77(2)  & 0.017(3) & 16 \\
Mn-Ni3 & 4.92(7)  & 0.0057(7) & 8 \\
\hline
\hline 
&&&\\

Ga K-edge & $k$ range: (2- 15)\AA$^{-1}$ & $k$-weight: 3 & $R$ range: (1-5)\AA\\
\hline
\hline
Bond Type & Bond Length  & $\sigma^2$ & Coordination \\
& R (\AA) & (\AA$^2$) & Number\\
\hline
Ga-Ni1 & 2.514(2) & 0.0081(3) & 8 \\
Ga-Mn1 & 2.734(9) & 0.011(1) & 4 \\
Ga-Mn2 & 3.20(2) & 0.040(4) & 2 \\
Ga-Ga1 & 3.90(2) & 0.017(8) & 4 \\
Ga-Ga2 & 4.24(2) & 0.013(3) & 8 \\
Ga-Ni2 & 4.74(2) & 0.015(2) & 16 \\
Ga-Ni3 & 4.92(2) & 0.008(1) & 8 \\
\hline
\hline 
  \end{tabular}
  \label{tab2}
\end{table}

\begin{figure}[h]
\epsfig{file=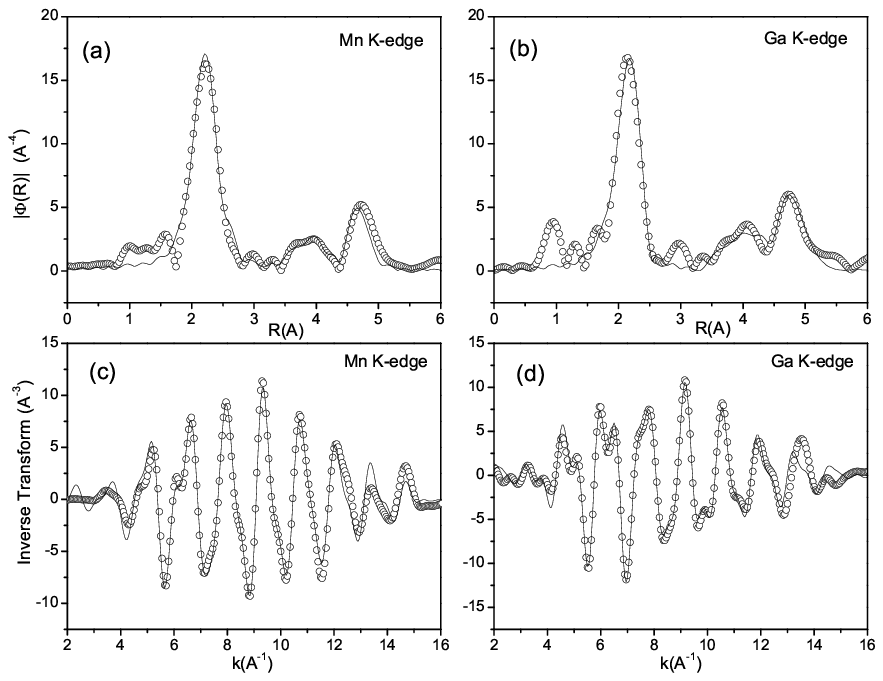, width=7cm, height=8cm}
\caption{\label{Ni216-xafs} Fit results for Mn and Ga K-edge EXAFS in Ni$_{2.16}$Mn$_{0.84}$Ga.The upper panel show the fit in R space and the lower panel show the $k^3$ weighted back transformed spectra. The circles represent the data and the solid line is the fit.}
\end{figure}

The starting model for EXAFS analysis for martensitic Ni$_{2.16}$Mn$_{0.84}$Ga was changed to an orthorhombic structure with space group Fmmm and lattice parameters a = b = 5.47\AA~ and c = 6.48\AA \cite{wedel}. The advantage of such an orthorhombic structural model is that only the lattice parameters change from that of the cubic $L2_1$ phase while the relative atomic coordinates remain unchanged. In the fits, the $\sigma^2$ values obtained for Ni$_2$MnGa served as starting parameters and $\delta R$ parameters were varied independently. The $\sigma^2$ values were subsequently refined to obtain better fitting.  Table\ref{tab2} presents the final fitted parameters and Fig.\ref{Ni216-xafs} presents the $R$-space and back transformed $k$-space fittings. The quality of the fit shows that the model adopted by us is quite satisfactory. 

\begin{figure}[h]
\epsfig{file=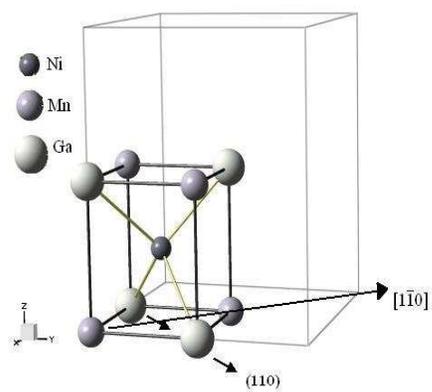, width=7cm, height=7cm}
\caption{\label{l10} The L1$_0$ sub-unit of the orthorhombic cell is shown. The arrows indicate the possible mechanism for movement of Ga atoms giving rise to modulations.}
\end{figure}

The important finding of the entire analysis is the change that occurs in the first shell of the central atom. The variation in the Mn-Ni and Ga-Ni bond distances obtained from the Mn and Ga edges in the two alloys reveal alot about the underlying martensitic transformations. From the Mn K-edge data analysis it is seen that the Mn-Ni bond distance increases from 2.523\AA~ to 2.532\AA~ in going from the austenitic to martensitic structure. However, from the Ga K-edge data analysis the Ga-Ni bond length does not follow the same trend. As seen from the Table\ref{tab1} and \ref{tab2}, Ga-Ni distance infact decreases by 0.007\AA~ from 2.521\AA~ to 2.514\AA. Furthermore, if one considers the difference between Mn-Ni and Ga-Ni bond distances in martensitic phase alone, this change is amplified to 0.018\AA. Both the central atoms, Mn and Ga can be viewed to be at the body centered position of a reduced tetragonal structure formed by Ni atoms. A non-uniformity in their bond distance with Ni of the order of 10$^{-2}$\AA ~ is unexpected and hints towards the microscopic changes influencing the formation of the macroscopic modulated phases. This directly points towards some sort of structural distortions in the orthorhombic cell. The disagreement in the value of bond lengths obtained from Ga K-edge analysis and from those obtained from Mn edge analysis based on the orthorhombic structural model can be understood if one considers the movement of Ga atom in the (110) plane along the [1$\bar 1$0]$_P$ direction. A modulated structure with Ga atom moved by 1\% from its crystallographic position of (0, 0, ${1\over{2}}$), in the cubic phase can fully explain the observed bond-lengths. This is summarized in pictorial form in Fig.\ref{l10}. 

It is this movement of Ga atoms that gives rise to modulations and the associated superstructures while the rigid Mn atoms forms an orthorhombic cell. It may be noted that the room temperature crystal structure of Ni$_{2.16}$Mn$_{0.84}$Ga has been reported in literature to be of 7M modulated type\cite{khov}.   

It appears from EXAFS analysis that when the sample undergoes martensitic transition, Ga and Ni atoms move closer to each other while the Mn atoms move away. A similar conclusion has also been drawn from the first principles calculation of electronic structure by Zayak et al\cite{zay-role}. Such a movement of atoms results in increase in $p-d$ hybridization at the Fermi level. Zayak et al\cite{zay1} have shown that the enhancement of interaction between Ga $p$ and Ni $d$ results in a dip in minority spin DOS at the Fermi level which is responsible for martensitic phenomena in Ni-Mn-Ga alloys. While the theoretical calculation\cite{zay-role} predicts a cooperative movement of all the atoms to be responsible for the modulated structure, our EXAFS analysis seems to show that atleast for the martensitic structures with $c/a > 1$ the movement of Ga atom alone can account for the same.

In summary, we have presented the local structural changes occuring in the crystal structure of Ni-Mn-Ga system on moving from austinite to martensite phase. The EXAFS study at Mn and Ga K-edges brings out very important aspects related to the changing $p-d$ hybridizations. The involvement of atomic displacements giving rise to modulated structures has been contemplated theoretically. However all such calculations considered movement of all the 3 constituent atoms from their crystallographic positions by different amounts\cite{pons, zay1}. The physical significance of the present study can be immediately appreciated as it clearly points out the movement of only one type of atom i.e. Ga to be responsible for a modulated structure.  

\acknowledgements
Authors gratefully acknowledge financial assistance from Department of Science and Technology, New Delhi, India and ICTP-Elettra, Trieste, Italy. Thanks are also due to Prof. G. Vlaic and Dr. Luca Olivi for help in EXAFS measurements and useful discussions. [P.A.B.] would like to thank Council for Scientific and Industrial Research, New Delhi for Senior Research Fellowship.

\end{document}